\RequirePackage[2020-02-02]{latexrelease}
\documentclass
[aps,preprint,showkeys,a4paper,tightenlines,superscriptaddress]{revtex4}%
\usepackage{eurosym}
\usepackage{amsfonts}
\usepackage{amsmath}
\usepackage{amssymb}
\usepackage{graphicx}
\usepackage{placeins}
\usepackage{subfig}
\usepackage{caption}
\usepackage{xcolor}%
\providecommand{\U}[1]{\protect\rule{.1in}{.1in}}
\providecommand{\U}[1]{\protect\rule{.1in}{.1in}}

\newcommand{\rev}[1]{\textcolor{black}{#1}}

\begin{document}

\title{
\rev{Finite deformations induce friction hysteresis in normal wavy contacts}
}
\author{M. Ceglie}
\affiliation{Department of Mechanics, Mathematics and Management, Polytechnic University of Bari, Via E. Orabona 4, Bari, 70125, Italy}
\author{G. Violano}
\email{guido.violano@poliba.it}
\affiliation{Department of Mechanics, Mathematics and Management, Polytechnic University of Bari, Via E. Orabona 4, Bari, 70125, Italy}
\author{L. Afferrante}
\affiliation{Department of Mechanics, Mathematics and Management, Polytechnic University of Bari, Via E. Orabona 4, Bari, 70125, Italy}

\keywords{nonlinear contact mechanics, friction, hyperelasticity, finite element method}

\author{N. Menga}
\affiliation{Department of Mechanics, Mathematics and Management, Polytechnic University of Bari, Via E. Orabona 4, Bari, 70125, Italy}
\keywords{nonlinear contact mechanics, friction, hysteresis, hyperelasticity, large deformations, finite element method}


\begin{abstract}
Since Hertz's pioneering work in 1882, contact mechanics traditionally grounds on linear elasticity, assuming small strains and displacements.
However, recent experiments clearly highlighted linear elasticity limitations in accurately predicting the contact behaviour of rubbers and elastomers, particularly during frictional slip, which is governed by geometric and material nonlinearity.

In this study, we investigate the basic scenario involving normal approach-retraction contact cycles between a wavy rigid indenter and a flat, deformable substrate. Both frictionless and frictional interfacial conditions are examined, considering finite strains, displacements, and nonlinear rheology. We developed a finite element model for this purpose and compared our numerical results with Westergaard's linear theory.

Our findings show that, even in frictionless conditions, the contact response is significantly influenced by geometric and material nonlinearity, particularly for wavy indenters with high aspect ratios, where normal-tangential stresses and displacements coupling emerges.
More importantly, interfacial friction in nonlinear elasticity leads to contact hysteresis (i.e., frictional energy dissipation) during \rev{normal loading-unloading} cycles. This behavior cannot be explained in a linear framework; therefore, most of the experiments reporting hysteresis are typically explained invoking other interfacial phenomena (e.g., adhesion, plasticity, or viscoelasticity). Here we present an additional suitable explanation relying on \rev{finite strains/displacements} with detailed peculiarities, such as vanishing pull-off force. Moreover, we also report an increase of hysteretic losses as for confined systems, stemming from the enhanced normal-tangential nonlinear coupling.

\end{abstract}
\maketitle


\section{Introduction}

Contact mechanics is one of the fundamental branches of tribology that examines the distribution of stresses, displacements, and gaps at the interface between deformable bodies \cite{maugis2013,johnson1987,barber2018,vakis2018}.
For several decades, the main contact mechanics studies have been based on the assumptions of small displacements and linear constitutive laws, with the aim of understanding the fundamental mechanisms governing the contact behaviour of solids. This approach has led to the development of very impacting theories \cite{bowden1939,westergaard1939,greenwood1966,persson2001,perssontosatti2001,carbone2011,Menga2014,joe2018} and numerical methods \cite{hyun2004,wriggers2006,campana2007,putignano2012,pastewka2014,yastrebov2015,muser2017,rey2017,afferrante2018} that have shed light on some effects such as surface roughness and interfacial adhesion. These insights extend to complex cases, including thin \cite{menga2016} and coated \cite{Menga2019coated} solids, as well as anisotropic surfaces \cite{Afferrante2019}.

Many studies indicate that linear theory approximation provides results aligning with experimental tests conducted on systems characterized by low aspect ratio and frictionless conditions \cite{bennett2017,mcghee2017,violanorate2}.

The presence of interfacial friction complicates the problem, even within the framework of linear theory, unless we assume (as is done in most classical contact mechanics studies) uncoupled normal and tangential elastic fields. This assumption implicitly requires similar materials (with a negligible second Dundurs' constant, $\beta$ \cite{dundurs1973}) and semi-infinite (half-space) bodies. However, under more practical conditions, these assumptions often do not hold, and friction becomes a critical factor. Since the late 1970s, it has been shown that local frictional slip alters the contact response between bodies of different materials, even in the simple case of Hertzian geometry \cite{Mossakovskii1963,Spence1975,Nowell1988etAl}, leading to a stiffer contact. Further solutions for partial slip problems can be found in \cite{Block2008}. Additionally, thin coatings applied to stiffer substrates (such as functional coatings and oxidation layers on metals) exhibit distinct responses under frictional sliding \cite{menga2019,menga2021,muller2023}, often resulting in significantly larger contact areas and an enhanced frictional response, known as 'geometric' friction.

Surfaces in many real-world applications are not "nominally" flat, as is common in numerous tribological interactions \cite{santeramo2023}. In such cases, large deformations and displacements introduce significant geometric nonlinearities. Furthermore, materials often exhibit nonlinear behaviour as seen in rolling tires \cite{gao2021} or human skin \cite{maiti2016}. At high loads and large deformations, the linear regime no longer applies, leading to plastic \cite{liu2014} and/or hyperelastic \cite{giannakopoulos2009,jobanputra2020} deformations.

Wriggers and colleagues have developed finite element (FE) methodologies that incorporate finite deformations into contact problems, including impact phenomena \cite{wriggers1990impact}, as well as frictionless \cite{taylor1905} and frictional interfaces \cite{wriggers2019}. While the integration of finite deformations into computational contact mechanics has largely remained a theoretical exercise for specialists in the field \cite{wriggers2006,laursen1993,pietrzak1999,kruse2015}, some studies have also investigated interfacial mechanisms such as the peeling of tapes/membranes within a finite deformation framework \cite{Molinari2008,Begley2013}. However, these studies have shown only quantitative differences compared to linear formulations \cite{Ceglie2022,Ceglie2024}, particularly in the case of thin deformable substrates \cite{Menga2020}.

Recent attention has shifted towards understanding how geometric and material nonlinearity affects contact mechanics. For example, experiments have shown that the contact area between an elastomeric sphere and a plane evolves from a circular shape under pure normal loading to a smaller, ellipsoidal area during macroscopic sliding \cite{savkoor1977,waters2010,sahli2018}, contradicting formal linear theory predictions \cite{MCD,MCD_corr}. Emerging hypotheses suggest that shear-induced reduction in the contact area is driven by the nonlinear, finite-deformation behaviour of the elastomer \cite{wang2020,mergel2021,lengiewicz2020,hess2020}, rather than being solely attributable to adhesion modulation \cite{savkoor1977,papangelo2019}. 
Another fascinating example is the phenomenon of Schallamach waves, first captured by Schallamach \cite{schallamach1971} in photographs of friction experiments involving rubber and glass. These waves arise from stick-slip instabilities on soft surfaces, appearing as cycles of wrinkles that propagate across the contact area in a wave-like motion. Their formation and movement are primarily driven by large deformation of rubber-like materials, involving frictional energy dissipation \cite{rand2006,fukahori2010}.

Even in the normal contact problem  between a rigid indenter and a deformable rubber-like material, energy dissipation is commonly observed in approach-retraction experiments. This phenomenon, known as contact hysteresis, is characterized by a mismatch between the approach and retraction paths and is typically attributed to adhesion \cite{chaudhury1993,Carbone2015}, viscoelasticity \cite{Carbone2022,Mandriota2024steady,Mandriota2024unsteady}, or plasticity \cite{li2021}.

In principle, contact hysteresis could arise also from interfacial friction, even in the absence of adhesive-viscoelastic phenomena or plastic deformations, when a certain degree of coupling exists between normal and tangential displacements. In linear theory, this coupling occurs only in the case of compressible materials\cite{Bentall1967,Nowell1988etAl} and/or thin layers\cite{menga2019,menga2021,muller2023}, while most elastomer contacts are considered incompressible and sufficiently thick to neglect such coupling. However, the role of geometric and material nonlinearity on approach-retraction hysteresis in these conditions has not been yet addressed.

To investigate the latter scenario, we address the long-standing problem of approach-retraction cycle of a wavy indenter into a deformable solid. We develop a finite element (FE) model able to \rev{simulate the contact behavior including nonlinear effects} arising from geometric factors, material properties, and \rev{interfacial} friction characteristics. 
To assess \rev{and highlight the role of nonlinearity, contact calculations are performed both in linear and nonlinear framework. Specifically, in the former case we adopt the well-known Westergaard's analytical solution \cite{westergaard1939}, both for frictionless and frictional contacts (uncoupled normal and tangential displacements); in the latter one, we assume finite strains/displacements with neo-Hookean hyperelastic behaviour, also introducing} Coulomb-Orowan interfacial friction \cite{Orowan1943}, which is typical of polymer contacts \cite{chateauminois2008,baillet1995}.

The results indicate that, even in the simplest case of frictionless contact, geometric and material nonlinearity lead to a contact response which \rev{differs} from the linear theory predictions. As expected, the discrepancy increases as the aspect ratio of the wavy indenter increases and, in turn, small-displacement and small-strain approximations are violated. Furthermore, accounting for friction and the real shape of the indenter (i.e., finite displacements - geometric effects) results in non-negligible contact hysteresis during loading and unloading cycles, both for semi-infinite deformable solids and, even more so, for thin layers.

Overall, our study provides new insights into the field of contact mechanics, highlighting the crucial role of contact nonlinearity. This also explains why theoretical models based on linear theory often fail to accurately capture experimental results, especially in the presence of friction.

\section{Problem formulation}

\begin{figure}[!ht]
\begin{center}
\includegraphics[width=0.5\textwidth]{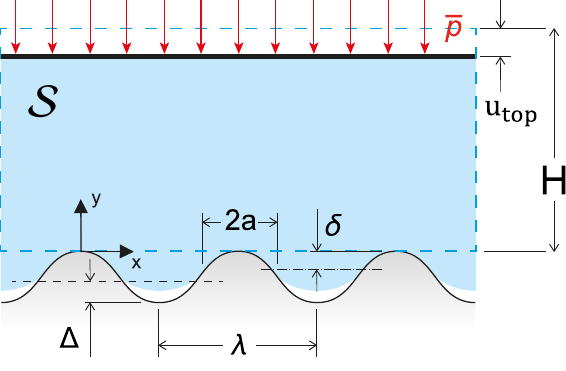}
\end{center}
\caption{A deformable solid $\mathcal{S}$ in contact with a rigid sinusoidal surface of amplitude $\Delta$ and wavelength $\lambda$. A remote uniform pressure $\Bar{p}$ is applied on the upper rigid slab bonded to the solid, leading to a contact mean penetration $\delta$ and a contact semi-width $a$.}%
\label{fig1}%
\end{figure}

The contact problem under investigation is illustrated in Fig. \ref{fig1}, where a flat deformable solid $\mathcal{S}$ of thickness $H$ backed by a rigid slab is pressed, with a mean remote pressure $\Bar{p}$, against a rigid sinusoidal indenter of wavelength $\lambda$ and amplitude $\Delta$.

\subsection{Finite displacement framework}

To account for geometric nonlinearity, our simulations are performed within a finite displacement framework, and contact quantities are defined relative to the current deformed configuration. Referring to Fig. \ref{fig2}, we define $(x,y)$ as the undeformed reference frame, centered on the crest of the rigid sinusoidal indenter, with $(\hat{i},\hat{j})$ as horizontal and vertical unit vectors, and indenter shape given as $r(x)=\Delta \cos(2\pi x/\lambda)-\Delta$. Displacement fields are always defined with reference to $(x,y)$. Consequently, for a generic point on the contact surface (i.e. $y=0$), the coordinates $(x',y')$ in the current deformed configuration \rev{at a generic time $t$} are determined by

\begin{equation}
	x'(x,t)=x+U_\textrm{x}(x,t) \text{and}  y'(x,t)=U_\textrm{y}(x,t)
    \label{eq1}
\end{equation}

where, $U_\textrm{x}$ and $U_\textrm{y}$ are the horizontal and vertical displacements fields, respectively. 
We also define the contact penetration $\delta$ as the mean normal displacement of the deformed surface \rev{calculated in the deformed configuration, i.e.,
\begin{equation}
    \delta =
    \frac{1}{\lambda} \int_{-\lambda/2}^{\lambda/2} U_\textrm{y} (x) \frac{dx'}{dx} dx
\end{equation}
where, according to Eq. (\ref{eq1}), $dx'/dx$ is the horizontal stretch of the local elemental length}.

Additionally, we define the local slope of the sinusoid $m(x)$ as
\begin{equation}
    m(x)=\tan{\phi(x)}=-\frac{2\pi \Delta}{\lambda} \sin(2\pi x/\lambda)
\end{equation}
where the local rotation $\phi$ is assumed to be positive when counterclockwise (as shown in Fig. \ref{fig2}), and the local unit tangential and normal vectors $(\hat{t},\hat{n})$ , respectively, as
\begin{equation}
    \begin{bmatrix}
       \hat{t}\\
       \hat{n}
    \end{bmatrix}=
    \begin{bmatrix}
       \cos{\phi} & \sin{\phi}\\
       -\sin{\phi} & \cos{\phi}
    \end{bmatrix}
    \begin{bmatrix}
       \hat{i}\\
       \hat{j}
    \end{bmatrix}
    \label{rot_matrix}
\end{equation}


Therefore, in the finite displacement scenario, the contact stress vector $\bold{s}$\rev{, corresponding to the Cauchy stress,} can be locally decomposed into the normal and tangential shear stresses, respectively
\begin{equation}
    p=\bold{s} \cdot \hat{n}   \text{and}     \tau=\bold{s} \cdot \hat{t}
    \label{p_tau}
\end{equation}

Similarly, to simplify the comparison between results under small and finite displacements conditions, we define the vertical $s_\textrm{y}$ and horizontal $s_\textrm{x}$ stresses as the projections of the (finite displacements) stress vector $\bold{s}$. Hence

\begin{equation}
    s_\textrm{x}(x')=(\bold{s} \cdot \hat{i})\sqrt{1+m^2}  \text{and}  s_\textrm{y}(x')=(\bold{s} \cdot \hat{j})\sqrt{1+m^2}
    \label{sy_sx}
\end{equation}
where the term $\sqrt{1+m^2}=dL'/dx'$ takes into account the increase in the local elemental length \rev{due to finite rotations (see Fig. \ref{fig2})}. Therefore, the remote mean pressure $\Bar{p}$ \rev{applied to balance the contact stresses is given by},
\begin{equation}
    \Bar{p}=\frac{1}{\lambda}\int_{-\lambda/2}^{\lambda/2}s_\textrm{y}dx'
    \label{p_bar}
\end{equation}

\subsection{Material and interfacial properties}

The solid $\mathcal{S}$ is assumed to be nearly incompressible, with a Poisson's ratio $\nu=0.49$. To highlight the effect of material nonlinearity, simulations are performed either for linear elastic and hyperelastic solids. In the latter case, neo-Hookean rheology is assumed; therefore, given the deformation gradient tensor $\bold{F}$, with $J=det(\bold{F})$, and the right Cauchy-Green deformation tensor $\bold{C}=\bold{F}^T\bold{F}$, the Neo-Hookean constitutive behaviour in nearly incompressible solids (i.e., $\nu\approx0.5$ and $J\approx1$) is described by the strain energy density function \cite{Bonet2008}
\begin{equation}
    W=\frac{\mu}{2}(\Bar{I}_1-3)+\frac{\kappa}{2}(J-1)^2
\label{eq:neo-hookean}
\end{equation}
where $\Bar{I}_1=tr(\bold{\bar{C}})$ is the first invariant of the distortional (isochoric) right Cauchy-Green deformation tensor $\bold{\bar{C}}=\bold{C}/J^{2/3}$, and $\mu$ and $\kappa$ are the shear and bulk moduli, respectively. For consistency with the linear elastic model, we set $\mu=E/[2(1+\nu)]$ and $\kappa=E/[3(1-2\nu)]$, where $E$ is Young's modulus and $\nu$ the Poisson's ratio.
Moreover, since in finite displacements analysis the equilibrium equations refer to the current configuration, the Cauchy stress tensor $\bold{\sigma}$ is required (i.e., local stresses in the current configuration). This can be calculated as
\begin{equation}
    \bold{\sigma}=\frac{2}{J}\bold{F}\cdot\frac{\partial W}{\partial \bold{C}}\cdot\bold{F}^T
\end{equation}

\begin{figure}[ht]
\begin{center}
\includegraphics[width=0.8\textwidth]{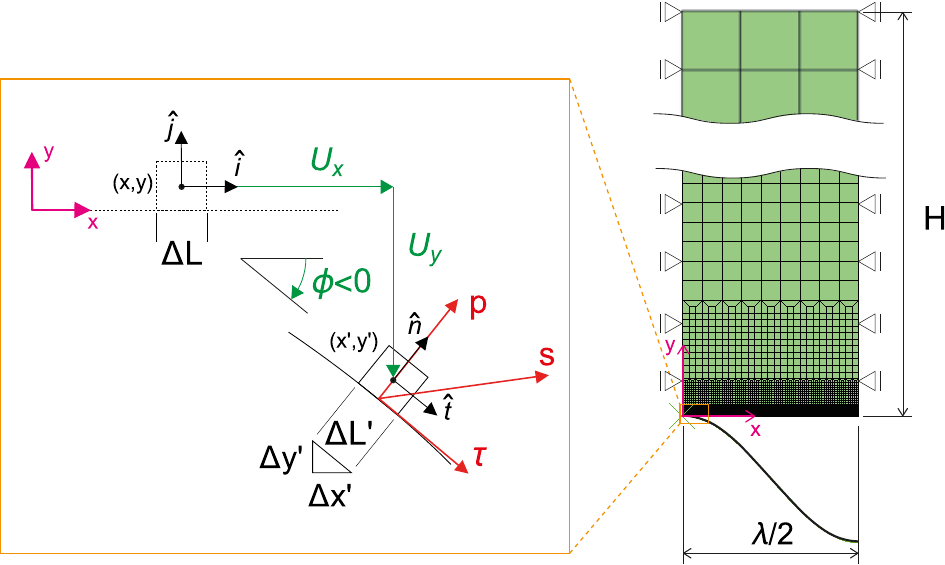}
\end{center}
\caption{FE mesh and details of the nonlinear geometric framework used in the analysis. Specifically, $\bold{s}$ is the local \rev{Cauchy} stress vector acting on the contact boundary of the solid $\mathcal{S}$ \rev{in the deformed configuration}, $(\hat{i},\hat{j})$ are the horizontal and vertical unit vectors in the undeformed configuration, while $(\hat{t},\hat{n})$ are the local unit vectors tangent and normal to the indenter surface, respectively (with local indenter slope \rev{$m$}). In the deformed configuration, the \rev{generic} interfacial element, with undeformed length $\Delta L$, is stretched to the length $\Delta L'$\rev{, is rotated by the angle $\phi$ (positive if counterclockwise)}, and is displaced in the horizontal and vertical directions by the quantities $U_\textrm{x}$ and $U_\textrm{y}$, respectively.}%
\label{fig2}%
\end{figure}

\rev{With reference to Fig.\ref{fig2},} the contact interface is assumed adhesiveless, \rev{i.e., $p \geq 0$ in Eq. (\ref{p_tau})}. In the tangential direction $\hat{t}$, both frictionless and frictional conditions are considered, in order to investigate how the finite strains/displacements may trigger frictional dissipation \rev{in nominally uncoupled conditions}, eventually affecting the overall contact behaviour. Based on experimental results, which indicate a minor effect of contact pressure on frictional shear stresses in polymers contacts \cite{chateauminois2008,nguyen2011}, we assume a velocity-independent Coulomb-Orowan friction law \cite{Orowan1943,baillet1995} with the threshold shear stress $\tau_{\textrm{sl}}$ for interfacial slip given by
\begin{equation}
      \tau_{\textrm{sl}}=\left\{
    \begin{array}{ll}
      \mu p, & \mbox{for  $\mu p<\tau_{\textrm{max}}$}\\
      \tau_{\textrm{max}}, & \mbox{for $\mu p \geq \tau_{\textrm{max}}$}
    \end{array}
    \right.
    \label{eq:friction_mod}
\end{equation}
where $\mu$ is the Coulomb friction coefficient, and $\tau_{\textrm{max}}$ represents the characteristic (or nominal) interface shear stress. Notably, the Coulomb-Orowan friction model also helps avoid convergence issues, which are likely to occur using a Tresca model (i.e., $\tau_{\textrm{sl}}=\tau_{\textrm{max}}$ for any value of the normal pressure $p$) \rev{due to the possible shear stresses discontinuity at the contact edges} \cite{lengiewicz2020}.

Additionally, being $\bold{v}_{\textrm{sl}}=v_{\textrm{sl}}\hat{t}$ the interfacial slip velocity, namely the local relative velocity between the solid surface and the indenter profile in the contact region, the frictional dissipation per unit time is given by
\begin{equation}
    \dot{W}_\textrm{F}=\int_{\partial \mathcal{S}} v_{\textrm{sl}} \tau_{\textrm{sl}} ds 
    \label{eq:friction_diss}
\end{equation}
where $\partial \mathcal{S}$ is the contact interface in deformed conditions.

\subsection{Finite Element Model}

The contact problem depicted in Fig. \ref{fig1} is solved with the aid of the commercial software SIMULIA Abaqus. Due to the symmetry and periodicity of the problem, the analysis is restricted to half of a single sinusoid. Consequently, horizontal displacements are assumed to vanish at the boundaries $x=0$ and $x=\lambda/2$\rev{, as shown in the right panel of Fig.\ref{fig2}}.

In classical contact mechanics, the half-plane assumption is often utilized to address contact problems, assuming that the contact area is small compared to the dimensions of the contacting bodies. To simulate this scenario, the height $H$ of the deformable solid $\mathcal{S}$ is taken to be sufficiently larger than the wavelength $\lambda$. Specifically, we verified that the half-plane approximation holds when $H \gtrsim 3\lambda$. Lower values of $H/\lambda$ are also considered in the second part of the paper to investigate the effects of finite thickness and mechanical confinement.

The solid $\mathcal{S}$ is meshed using linear plane strain quadrilateral elements with the reduced integration scheme to prevent shear locking, along with a hybrid formulation to accurately capture the nearly incompressible material behaviour. A regular mapped mesh is employed, featuring finer discretization at the contact interface where the element size is $\Delta x = \lambda/729$, and a coarser mesh in the upper part of the solid. This approach optimizes computational efficiency. A preliminary mesh convergence test has been conducted.
The rigid indenter is modeled using an analytical rigid surface, where displacement and force can be assigned to a single master node. Simulations are conducted on sinusoidal profiles with an aspect ratio in the range $0.1\leq\Delta/\lambda\leq0.5$, since in most engineering surfaces, the asperity width is considerably larger than its height \cite{Scaraggi2016}.
Normal contact is modeled using a hard contact pressure-overclosure relationship, and the augmented Lagrangian method is adopted to enforce the contact constraints. 
\rev{A penalty formulation is adopted to solve the frictional contact problem; therefore, the slope of the frictional stress versus total slip relationship (often referred to as the sticking stiffness) is finite even when sticking occurs, and the total slip is composed of an allowable “elastic” slip and an “inelastic” slip. Only the latter term corresponds to the actual slip locally occurring where $\tau=\tau_{\textrm{sl}}$ and, in turn, is associated with frictional dissipation, i.e. the term $v_\textrm{sl}$ in Eq. (\ref{eq:friction_diss}).}
Indentation tests are performed \rev{as static analysis,} under normal force-controlled conditions. The contact between the bodies is enforced by applying a uniform pressure $\Bar{p}$ on the rigid slab bonded to the upper boundary of the solid $\mathcal{S}$. During a loading-unloading cycle, the pressure $\Bar{p}$ gradually ramps from zero to the target value $\Bar{p}_{\textrm{max}}$ and then back to zero. 


Finally, to address instabilities common in nonlinear static problems, we employ a numerical stabilization methodology based on the addition of volume-proportional damping. The stabilization parameters are adjusted following preliminary tests to ensure model convergence and avoid inaccurate results. In more detail, by evaluating the impact of the stabilization in terms of dissipated energy, we find that it remains below $10^{-7}\%$ of the internal elastic energy and below $10^{-4}\%$ of the frictional energy dissipation throughout the entire process.


\subsection{The Westergaard's model}
The contact problem described above has been extensively approached in a fully linear framework by means of analytical and numerical tools. In 1939, Westergaard \cite{westergaard1939} derived an analytical closed-form solution for the frictionless contact problem of a linear elastic half-plane (i.e., $H/\lambda \rightarrow \infty$) and a wavy rigid surface. He found that the applied \rev{mean} pressure $\Bar{p}$ can be related to the contact size $2a$ by
\begin{equation}
    \Bar{p}=\frac{E^*\lambda}{4\pi R}\sin^2{\psi_\textrm{a}},
\end{equation}
where $\psi_\textrm{a}=\pi a/\lambda$, $E^*=E/(1-\nu^2)$, and $R=\lambda^2/(4\pi^2\Delta)$ is \rev{the radius of} curvature at the wave crest.

The pressure distribution within the contact area is instead given by
\begin{equation}
    p(x)=\frac{E^*\lambda}{2\pi R}\cos{\psi}\sqrt{\sin^2{\psi_\textrm{a}}-\sin^2{\psi}}
\end{equation}
with $\psi=\pi x/\lambda$.

The above relations hold true in the small displacements approximation, namely when the indenter slope is sufficiently small to assume that the normal pressure $p(x)$ is vertical\rev{, i.e. $\hat{n}=\hat{j}$ along the whole contact interface.}

\section{Results} \label{sec_results}

\begin{figure}[ht]
\begin{center}
\includegraphics[width=0.8\textwidth]{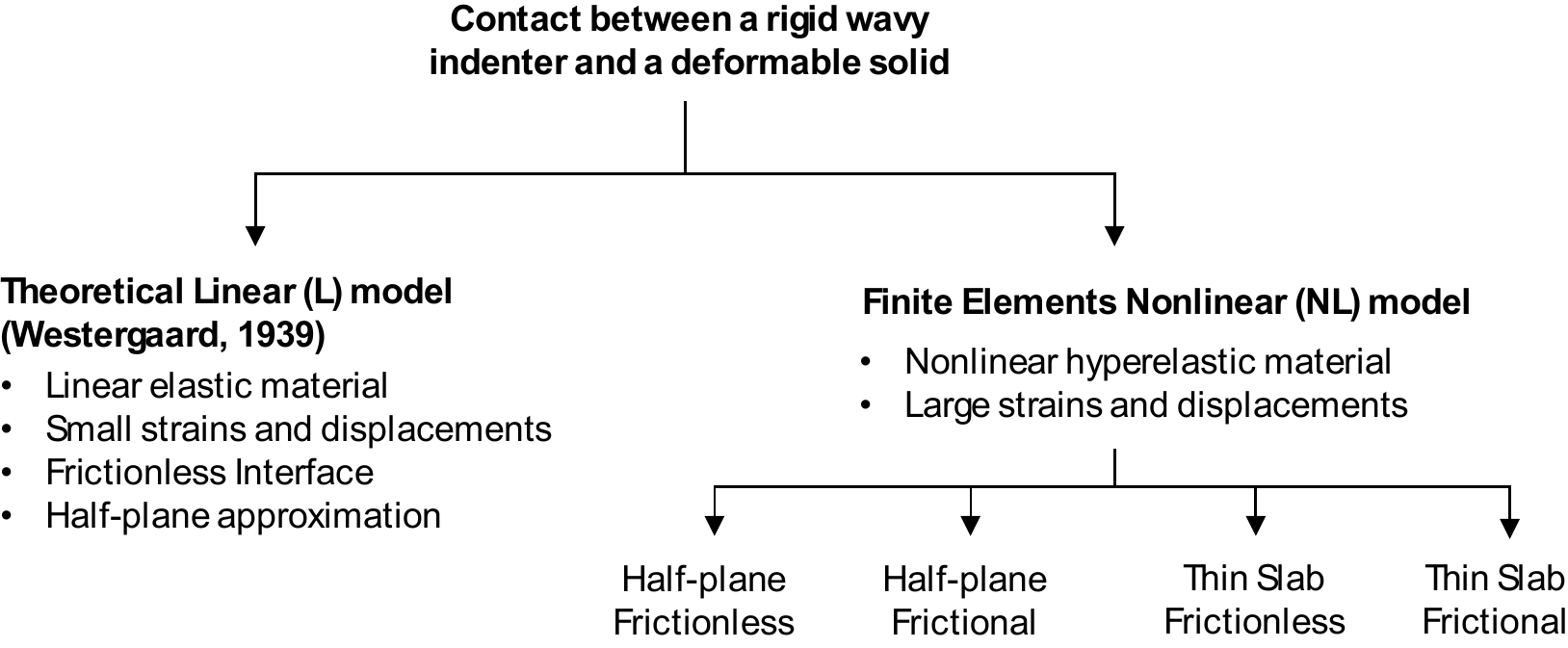}
\end{center}
\caption{\rev{The models used to solve the contact problem: the Westergaard's linear model (L), derived within the framework of linear elasticity, and the finite elements nonlinear model (NL), which incorporates finite strains/displacements formulation for a hyperelastic noe-Hookean material. The key assumptions and features of each model are summarized in the chart.}}%
\label{figCHART}%
\end{figure}

The scope of this study is to investigate the nonlinear effects arising in both frictionless and frictional normal contacts between a deformable rubber-like solid and a rigid wavy countersurface. According to the chart reported in Fig. \ref{figCHART}, as a reference linear solution for semi-infinite solids (half-plane), we adopt the Westergaard's analytical solution \cite{westergaard1939}, referred to as the linear (L) model. Importantly, since linear elasticity predicts that contacts of incompressible half-planes exhibit uncoupled normal and tangential elastic fields, Westergaard's solution still holds true for frictional contacts under pure normal loading.

To overcome the limitations of the L model, we developed a finite element model able to account for \rev{the geometric and material nonlinearity, and the finite dimensions of the deformable solid. Indeed, the NL model incorporates finite strains/displacements and the hyperelastic neo-Hookean rheology given in Eq. (\ref{eq:neo-hookean})}. 
\rev{In what follows, we perform a systematic comparison between the results of L and NL models in order to highlight the role of nonlinearity in the contact response.} This eventually results in an effective coupling between normal and tangential elastic fields, even under frictionless conditions. Since friction is ineffective for the L model, higher discrepancy is expected in frictional conditions\rev{, also involving frictional energy dissipation and hysteresis in normal loading-unloading cycles for the NL model}.


The results are presented for $E=1$ MPa, $\nu=0.49$, $\lambda=5$ mm, $\Delta/\lambda=0.1, 0.18, 0.3, 0.5$. The half-plane approximation is simulated by assuming $H/\lambda=5$, while finite thickness scenarios are explored for $0.25\leq H/\lambda\leq5$.

\rev{The effect of friction is also investigated. Since in Coulomb-Orowan friction regime, the overall friction mainly depends on the value of $\tau_{\textrm{max}}$ due to shear stresses saturation in the contact, simulations are carried out for different values of $\tau_{\textrm{max}}$, varying in the range 0.01÷0.2 MPa. Conversely, calculations have shown that reducing $\mu$ by an order of magnitude results in less than a 1\% variation in both the contact area and frictional dissipation; therefore, we only focus on the realistic case of $\mu=1$.}


In presenting the results, we shall refer to the following dimensionless quantities: contact radius $a/\lambda$, contact pressure $p/E^*$, characteristic shear stress $\tau_{\textrm{max}}/E^*$, shear stress $\tau/\tau_{\textrm{sl}}$, frictional dissipation $W_\textrm{F}/(E^*\lambda^2)$, contact profile coordinates $x/\lambda$ and $y/\Delta$, penetration $\delta/\Delta$, displacements $U/\Delta$. The results are organized into three subsections: first, geometric and material nonlinear effects are examined for the half-plane approximation under frictionless interface conditions; next, the influence of friction on contact quantities is analysed, both for a simple loading (approach) phase and for a full loading-unloading cycle; finally, finite-size effects in the NL model due to 'thin' solids confinement are investigated under both frictionless and frictional conditions.

\subsection{\rev{Frictionless contacts}}
\begin{figure}[!ht]
\begin{center}
\includegraphics[width=1\textwidth]{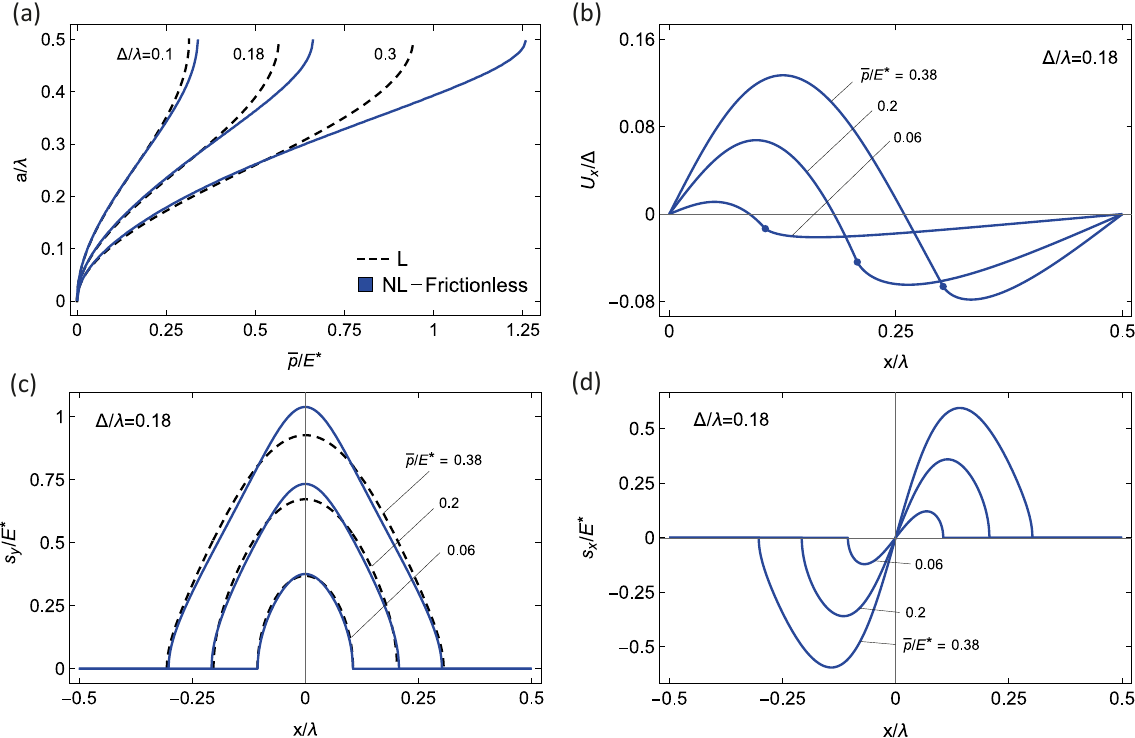}
\end{center}
\caption{\rev{The normal loading phase in frictionless interfacial conditions for the nonlinear (NL) FE model (blue curves). Analytical linear (L) solution (Westergaard) is shown in dashed black for comparison.} (a) Dimensionless contact semi-width $a/\lambda$ versus dimensionless \rev{mean} applied pressure $\Bar{p}/E^*$ during loading (approach) phase, for different values of the aspect ratio $\Delta/\lambda$; dimensionless (b) horizontal displacement $U_\textrm{x}/\Delta$ (circles indicate the contact edge), (c) vertical stresses distribution $s_\textrm{y}/E^*$, and (d) horizontal stress distribution $s_\textrm{x}/E^*$, for different values of the dimensionless applied pressure $\Bar{p}/E^*$. 
}
\label{fig4}%
\end{figure}

\rev{In this section, we compare the results of the linear elastic theory (L - black dashed lines) and the nonlinear finite element model (NL - blue lines) assuming a frictionless behavior at the solid-indenter interface. Specifically,}
Figure \ref{fig4}a shows the dimensionless contact area $a/\lambda$ as a function of the dimensionless mean applied pressure $\Bar{p}/E^*$, for different values of the indenter aspect ratio $\Delta/\lambda$. \rev{As expected, in the limit of nominally smooth indenters with $\Delta/\lambda\ll1$, the linear and nonlinear results overlap up to full-contact. Increasing the value of $\Delta/\lambda$, the linear prediction is still in qualitative agreement with NL results only at relatively low values of $\Bar{p}/E^*$. More in detail, the higher the aspect ratio, the lower the pressure threshold for linear prediction failure.}

Indeed, Westergaard's predictions (L) overestimate the contact area at high applied pressures compared to nonlinear calculations (NL), \rev{suggesting that nonlinearity entails an overall stiffening of the contact. A possible explanation is that,} at large pressures, most of the solid experiences compressive stresses \rev{which, in neo-Hookean rheology, are associated to a stiffer response than the corresponding linear material.} As a result, higher loads are required to achieve full contact in the NL model compared to the linear solution.

This is also clearly shown in Figs. \ref{fig4}c and \ref{fig4}d, where the vertical $s_\textrm{y}/E^*$ (Fig. \ref{fig4}c) and the horizontal $s_\textrm{x}/E^*$ (Fig. \ref{fig4}d) stress components are plotted for different values of the dimensionless applied pressure $\Bar{p}/E^*$\rev{, according to Eq.(\ref{sy_sx})}. We note the classical antisymmetric distribution of $s_\textrm{x}/E^*$ and, more importantly, the higher peak value in the vertical stress component $s_\textrm{y}$ observed for \rev{the NL model at relatively high mean pressure values (i.e., $\Bar{p}/E^*\gtrapprox0.2$)}, due to the stiffer compressive response of the Neo-Hookean material.

\rev{A remarkable effect of finite displacements (i.e., geometric nonlinearity) is the occurrence of non-vanishing values of $s_\textrm{x}$, even in frictionless conditions. This is} related to the finite local contact slope $m(x)$. \rev{Indeed, combining Eqs. (\ref{rot_matrix},\ref{sy_sx})} we locally have $s_\textrm{x}/s_\textrm{y} = -m$ and, since the maximum slope is $m_{\textrm{max}} = -2\pi\Delta/\lambda$, it follows that $s_\textrm{x}/s_\textrm{y}$ \rev{can locally approach the unit value,} for $\Delta/\lambda \approx 0.1$. However, while the latter aspect ratio value is typically admitted in linear analysis, \rev{the corresponding (L) frictionless prediction ($s_\textrm{y}=|\textbf{s}|$ and $s_\textrm{x}=0$) is far from reality. This discrepancy also applies to the horizontal displacement field $U_\textrm{x}$, which} are expected to vanish under frictionless linear conditions, unless other sources of \rev{in-plane/out-of-plane} coupling exist (e.g., material compressibility and/or solid confinement, none of which are considered in this case).
\rev{Conversely, since geometric nonlinearity is accounted for in the NL model}, coupling between normal and tangential displacements arises leading to non-zero horizontal displacements $U_\textrm{x}$, as clearly shown in Fig. \ref{fig4}b.

\subsection{\rev{Frictional contacts}}
\rev{Since finite strains/displacements formulation show significant discrepancies in contact response compared to linear predictions even in the simpler case of frictionless interfaces, here we extend the study to the frictional case. Indeed, we expect the observed coupling between normal and tangential displacements to be exacerbated by frictional shear stresses, which are modelled according to the Coulomb-Orowan friction law \cite{Orowan1943,baillet1995}, as described in Eq. (\ref{eq:friction_mod}). Firstly, we focus on a simple loading phase, then we consider a full loading-unloading cycle. 
As discussed at the start of Sec. \ref{sec_results}, Westergaard's linear solution (L) still applies to frictional conditions under purely normal loading and is reported for comparison.}

\begin{figure}[!ht]
\begin{center}
\includegraphics[width=1\textwidth]{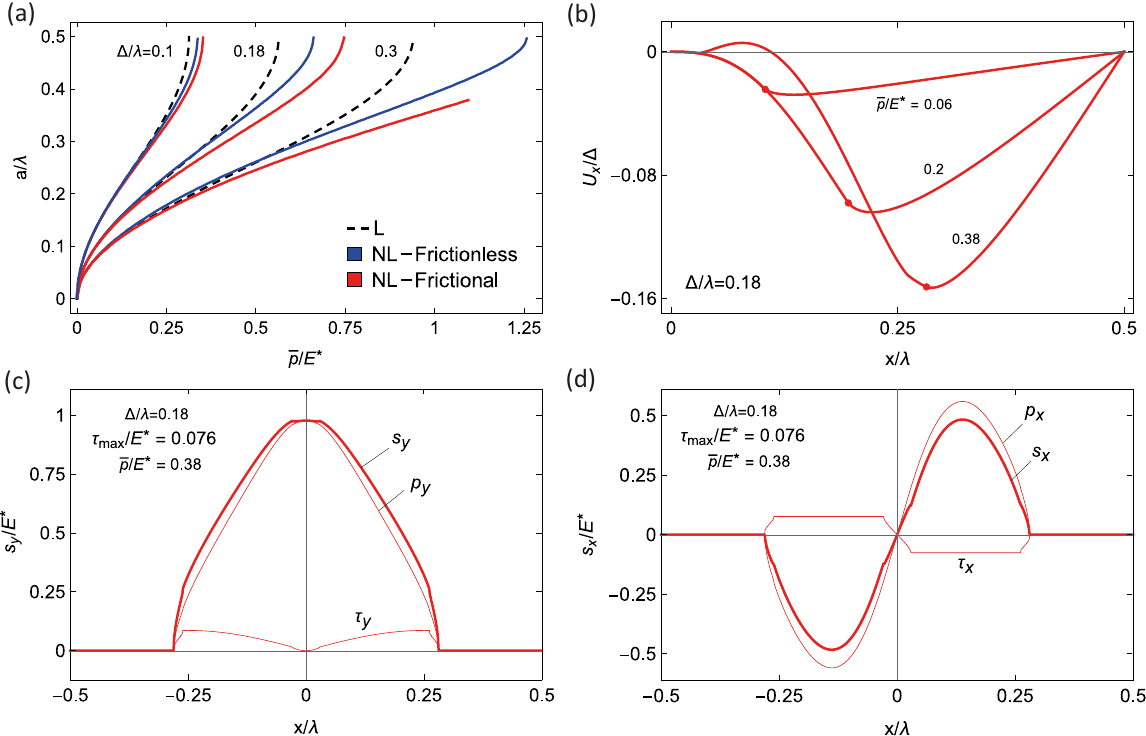}
\end{center}
\caption{\rev{The normal loading phase in frictionless (blue curves) and frictional (red curves) interfacial conditions for the nonlinear (NL) FE model. Analytical linear (L) solution (Westergaard) is shown in dashed black for comparison.}
(a) Dimensionless contact semi-width $a/\lambda$ versus dimensionless applied pressure $\Bar{p}/E^*$ during loading (approach) phase, for different values of aspect ratio $\Delta/\lambda$. A comparison between frictionless and frictional contact is proposed; (b) Dimensionless horizontal displacement at the contact interface $U_\textrm{x}/\Delta$ (circles indicate the contact edges) for different applied pressure $\Bar{p}/E^*$; (c) Dimensionless vertical stress distribution $s_\textrm{y}/E^*$, \rev{decomposed in} the pressure ($p_{\textrm{y}}$) and shear stress ($\tau_{\textrm{y}}$) \rev{contributions}; (d) Dimensionless horizontal stress distribution $s_\textrm{x}/E^*$ \rev{decomposed in} the pressure ($p_{\textrm{x}}$) and shear stress ($\tau_{\textrm{x}}$) \rev{contributions}. 
}
\label{fig5}%
\end{figure}

\rev{Figures \ref{fig5} refer to the case of a simple indentation between the sinusoidal indenter and the deformable solid.
More in detail, Fig. \ref{fig5}a shows a straightforward comparison between NL frictionless and frictional results showing that, in the latter case, frictional shear stresses opposing the relative interfacial slip stiffen the contact response. Indeed, especially at relatively high values of $\Bar{p}/E^*$, the contact area in frictional conditions is smaller compared to the frictionless case. 
Moreover, as long as friction prevents interfacial tangential slip (i.e., at relatively low mean pressure), the horizontal displacements $U_\textrm{x}$ are inward due to finite rotation of the adhering surface elements, as shown in figure \ref{fig5}b, in contrast to what is observed in frictionless conditions ($U_\textrm{x}$ pointing outward in the contact area). Increasing the applied mean pressure leads to outward slip in a portion of the contact area, and outward values of $U_\textrm{x}$ can also be locally observed (as for $\Bar{p}/E^*=0.38$ in Fig. \ref{fig5}b). These arguments indicate that inward frictional shear stresses are expected during loading, as indeed confirmed in Fig. \ref{fig5}d where, combining Eqs. (\ref{p_tau},\ref{sy_sx}), the horizontal projection $s_\textrm{x}$ of the stress vector is decomposed into pressure and frictional shear stress contribution, $p_x$ and $\tau_\textrm{x}$ respectively. Furthermore, this also helps in qualitatively explaining the stiffer response of frictional contacts; indeed, as shown in Fig. \ref{fig5}c, due to finite displacements, the vertical component $\tau_\textrm{y}$ of frictional shear stresses contributes to $s_\textrm{y}$ and, in turn, to balance the applied normal pressure $\Bar{p}$ (see Eq. (\ref{p_bar})). Since in frictionless conditions $\tau_\textrm{y}=0$, we have $s_\textrm{y}=p_\textrm{y}$ and a softer contact response.}

The high local contact stresses and large deformations near the asperity crests \rev{in frictional conditions} led to solver instability, causing the curve to truncate at very high pressures and aspect ratios. The results within the range shown, however, are valid and support the overall conclusions.


\begin{figure}[!ht]
\begin{center}
\includegraphics[width=\textwidth]{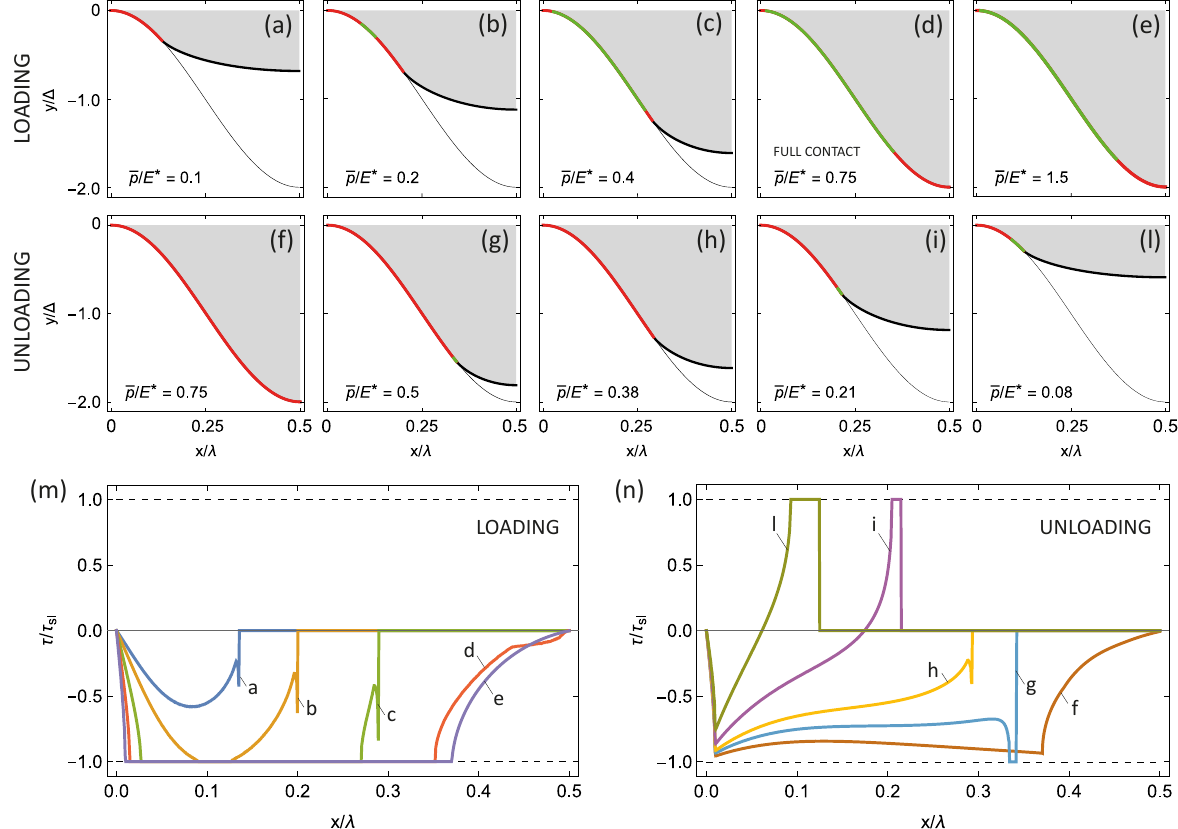}
\end{center}
\caption{\rev{The normal loading-unloading cycle for the FE nonlinear model (NL) in frictional conditions}. The dimensionless contact shape during a full loading (a-e) - unloading (f-l) cycle with dimensionless target mean pressure $\Bar{p}_{\textrm{max}}/E^*=1.5$. Red and green colours refer to contact area regions in stuck and slip conditions, respectively. 
The normalized contact shear stresses distribution $\tau/\tau_{\textrm{sl}}$ during the loading (m) and unloading (n) processes corresponding to Figs. a-l.
Results refer to $\Delta/\lambda=0.18$, $\tau_{\textrm{max}}/E^*=0.76$ and $\Bar{p}_{\textrm{max}}/E^*=1.5$
}%
\label{fig6}%
\end{figure}

Fig. \ref{fig6} shows the deformed profile during a complete loading-unloading cycle (Figs. \ref{fig6}a-l) and the corresponding distribution of the contact shear stress $\tau/\tau_{\textrm{sl}}$ within the contact area (Figs. \ref{fig6}m,n). Red and green colours indicate contact regions in stuck and slip conditions, respectively. 
Let us first focus on the loading phase. At sufficiently low values of $\Bar{p}/E^*$ (Fig. \ref{fig6}a), no slip occurs within the contact zone, as the interfacial shear stresses do not exceed the critical value $\tau_{\textrm{sl}}$ across the entire contact area ($|\tau|/\tau_{\textrm{sl}} \leq 1$).
As the pressure increases, outward slip begins at the interface, initially in a narrow region at the center of the contact semi-width (Fig. \ref{fig6}b), and then expands over a wider zone (Fig. \ref{fig6}c) until full contact is achieved (Fig. \ref{fig6}d). 
The onset of outward slip at the center of the contact semi-width, as shown in Fig. 6b, occurs due to the increased local slope and curvature of the indenter in this region. Indeed, in the framework of geometric nonlinearity, since equilibrium depends on the deformed interface geometry, the tangential stresses increase rapidly (eventually causing slip) where the indenter's slope is steeper, as they partially contribute to balance the (vertical) remote load. This is a non-trivial result strictly related to nonlinear analysis, while linear elasticity partial slip problems usually predict slip initiation at the contact edges due to shear stresses concentration and successive propagation toward the inner contact regions \cite{Spence1975}.
Interestingly, since the material is nearly-incompressible ($\nu=0.49$), once full contact is achieved (Fig. \ref{fig6}d), a further increase in the applied pressure still causes a displacement of the upper boundary of the solid (i.e., the rigid slab), resulting in changes in the overall displacements fields and additional slip at the interface (Fig. \ref{fig6}e).

Now, consider the unloading process. In the initial phase, the applied pressure decreases across the entire contact zone while full contact is maintained, and no slip occurs, unlike during the loading phase (Fig. \ref{fig6}f).

As the contact area recedes, slip initiates at the edge of the contact zone. In this case, as the externally applied pressure decreases, the overall contact pressure reduces, but the corresponding reduction in the contact area leads to a local concentration of contact stresses near the edge, resulting in $\tau>\tau_{\textrm{sl}}$ in this region.

As the applied pressure decreases further, shear stress near the contact edge also decreases, eventually falling below $\tau_{\textrm{sl}}$, leading to a recession of the contact without slip (Fig. \ref{fig6}h). As the unloading process continues, the contact shear stress reverses and concentrates at the contact edge. At this point, slip occurs inward, i.e., in the opposite direction compared to the loading process (Figs. \ref{fig6}i,l).

\begin{figure}[!ht]
\begin{center}
\includegraphics[width=\textwidth]{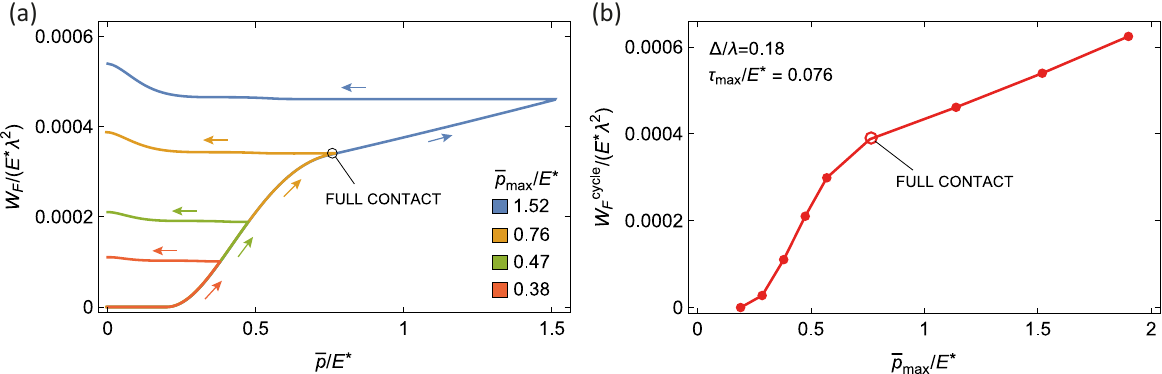}
\end{center}
\caption{\rev{The normal loading-unloading cycle for the FE nonlinear model (NL) in frictional conditions}. (a) The dimensionless frictional energy dissipation $W_\textrm{F}/(E^*\lambda^2)$ as a function of the dimensionless applied pressure $\Bar{p}/E^*$ during loading-unloading cycles. Notably, the effect of different values of dimensionless target applied pressure $\Bar{p}_{\textrm{max}}/E^*$ is shown. (b) The total dimensionless energy dissipated in a full cycle $W_\textrm{F}^{\textrm{cycle}}/(E^*\lambda^2)$ as a function of the dimensionless remote target pressure $\Bar{p}_{\textrm{max}}/E$.
Results are obtained for $\Delta/\lambda=0.18$ and $\tau_{\textrm{max}}/E^*=0.76$.}
\label{fig7}%
\end{figure}

Figure \ref{fig7}a illustrates the dimensionless energy dissipated in frictional relative slip, $W_\textrm{F}/(E^*\lambda^2)$, during the loading-unloading cycle path, plotted against the actual applied pressure, $\Bar{p}/E^*$. The results are presented for indentations performed at different remote target mean pressures, $\Bar{p}_{\textrm{max}}/E^*$. Similarly, Fig. \ref{fig7}b shows the dimensionless total energy dissipated over the whole cycle, $W_\textrm{F}^{\textrm{cycle}}/(E^*\lambda^2)$. $W_\textrm{F}^{\textrm{cycle}}$ is obtained by integrating Eq. (\ref{eq:friction_diss}) over the complete cycle. Alternatively, it can be calculated as the work done during loading-unloading by the applied remote pressure $\Bar{p}$, namely $W_\textrm{F}^{\textrm{cycle}}=\int_{\textrm{cycle}}\Bar{p}du_{\textrm{top}}$, where $u_{\textrm{top}}$ is the normal displacement of the rigid slab bonded to the upper solid boundary. 
Both figures clearly show that increasing the pressure beyond the full-contact threshold leads to greater energy dissipation, consistently with the displacement rearrangements discussed earlier. 
More specifically, \rev{Fig. \ref{fig7}a clearly shows} that dissipation primarily occurs during the loading phase, and the amount of energy dissipated during slip under full-contact conditions increases linearly with the applied pressure. Consequently, once full contact is established, the total energy dissipated over a complete cycle becomes almost linearly dependent on the target applied pressure\rev{, as shown in Fig. \ref{fig7}b}.

\begin{figure}[!ht]
\begin{center}
\includegraphics[width=\textwidth]{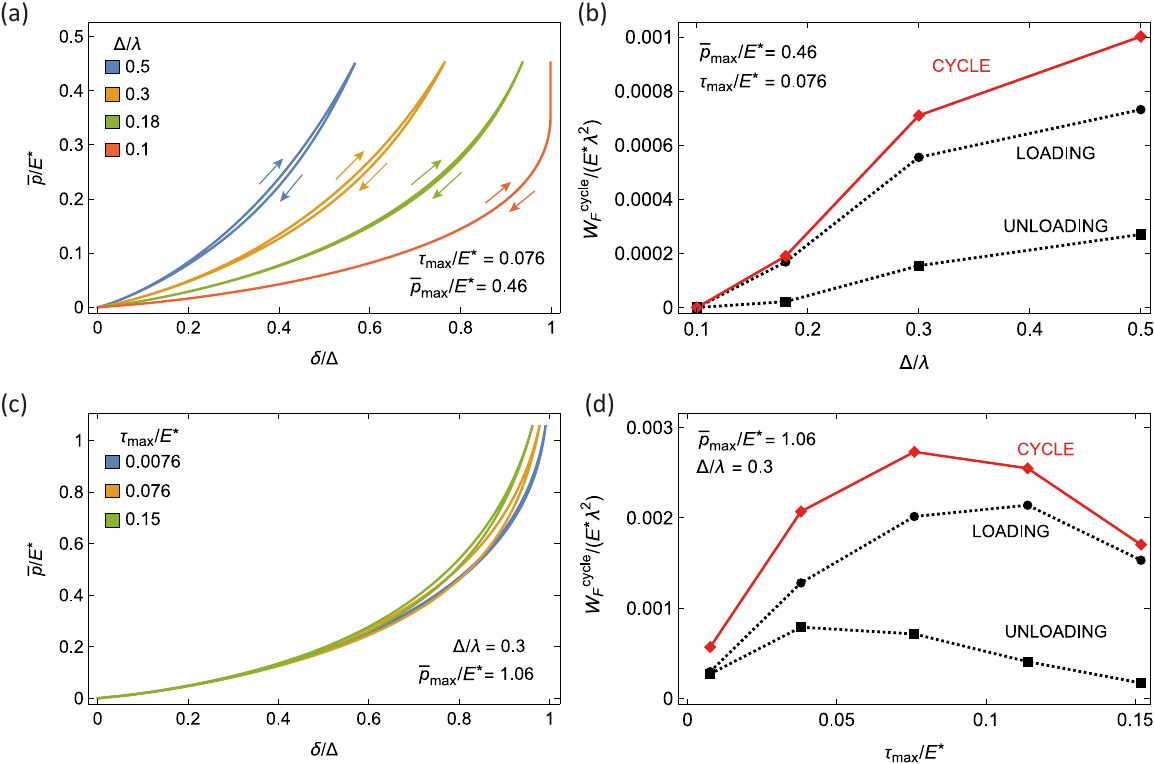}
\end{center}
\caption{
Frictional loading-unloading behaviour of the nonlinear model (NL) for different geometric and frictional parameters. Specifically, the dimensionless applied pressure $\Bar{p}/E^*$ as a function of the dimensionless penetration $\delta/\Delta$ for different asperity ratios $\Delta/\lambda$ (a), and dimensionless characteristic interface shear stress $\tau_{\textrm{max}}/E^*$ (c).
The dimensionless energy dissipated in a full indentation cycle $W_\textrm{F}^{\textrm{cycle}}/(E^*\lambda^2)$ as a function of $\Delta/\lambda$ (b) and $\tau_{\textrm{max}}/E^*$ (d) (red line); loading and unloading contributions (dotted lines) are also shown.}
\label{fig8}%
\end{figure}

Figure \ref{fig8}a shows the influence of the indenter aspect ratio on the load-displacement curves during a loading-unloading cycle in the presence of friction, while Fig. \ref{fig8}b illustrates the dimensionless total energy dissipated over a complete cycle as a function of the aspect ratio $\Delta/\lambda$. The results are presented \rev{referring to} a dimensionless target mean pressure of $\Bar{p}_{\textrm{max}}/E^{*} = 0.46$.

Smooth indenters (i.e., low values of $\Delta/\lambda$) do not induce significant finite strain or displacement effects, and the loading and unloading curves follow the same path. In this case, frictional dissipation vanishes as the whole contact is in stuck conditions, making the loading process fully reversible. As $\Delta/\lambda$ increases, frictional hysteresis appears. Specifically, Fig. \ref{fig8}a shows that, for a given contact penetration, the applied pressure during the unloading phase is lower than that during the loading phase, due to the need to overcome residual friction. As shown in Fig. \ref{fig8}b, energy dissipation increases with the indenter aspect ratio, driven by large deformations and finite displacements that trigger frictional slip.

Notably, loading-unloading hysteresis is typically attributed to different interfacial phenomena, such as elastic adhesion in smooth (chemically activated) contacts \cite{chaudhury1993} and rough contacts \cite{Carbone2015}, viscoelastic adhesion \cite{dalvi2019,violano2021,Carbone2022,Mandriota2024steady,Mandriota2024unsteady}, and/or plasticity \cite{li2021}, none of which are included in the present model. However, while adhesion-related phenomena generally lead to a jump out of contact with a finite contact area and pull-off force, we find that frictional hysteresis triggered by finite strains and displacements results in a smooth reduction of the contact area to zero, with a vanishing pull-off force.

Figures \ref{fig8}c and \ref{fig8}d focus on the effect of the characteristic shear stress $\tau_{\textrm{max}}$. Figure \ref{fig8}c presents the load-displacement curves for different values of $\tau_{\textrm{max}}/E^*$, while Fig. \ref{fig8}d shows the dimensionless total energy dissipated over a complete cycle, $W_\textrm{F}^{\textrm{cycle}}/(E^*\lambda^2)$, as a function of $\tau_{\textrm{max}}/E^*$.

As $\tau_{\textrm{max}}/E^*$ increases, higher stresses develop within the solid, which, combined with nonlinear rheology, result in a stiffer contact response, as seen in Fig. \ref{fig8}c. On the other hand, Fig. \ref{fig8}d reveals the presence of a value of $\tau_{\textrm{max}}/E^*$ that corresponds to a maximum in frictional energy dissipation per cycle. At low shear stress levels, although local slip begins at lower pressures and affects a larger portion of the contact area, the overall contribution to energy dissipation remains minimal. In contrast, at high shear stress levels, only a small portion of the contact area experiences slip, again leading to low-energy dissipation.

Our findings on hysteresis and energy dissipation, driven by geometric and material nonlinearity, have practical relevance in systems involving relatively soft elastomeric contacts. For example, understanding these mechanisms can enhance the functionality (energy dissipation) and the durability (wear) of tires, seals, and soft robotic grippers whose interfaces typically experience high friction and deformations, due to the aspect ratio of typical countersurfaces such as road asphalt \cite{lorenz2013}.

\subsection{Finite-size effects in confined contact}
\rev{As known, the half-plane approximation is no longer appropriate} when the size of the contact area approaches the \rev{thickness} of the contacting bodies, especially in compliant solids undergoing large deformations \cite{violano2024}. Recent studies have shown that contact confinement can result in a \rev{degree} of coupling between normal and tangential displacement fields, even within \rev{the framework of linear elasticity} \cite{menga2019,menga2021,muller2023}. \rev{Since previously discussed results showed that such a kind of coupling is significantly affected by finite strains/displacements even in nominally uncoupled conditions (i.e., incompressible half-plane with $H/\lambda \gg 1$), in this section we focus on thin solids ($H/\lambda \ll 1$) to explore the role of nonlinearity on the contact behavior and frictional dissipation.}

\begin{figure}[!ht]
\begin{center}
\includegraphics[width=\textwidth]{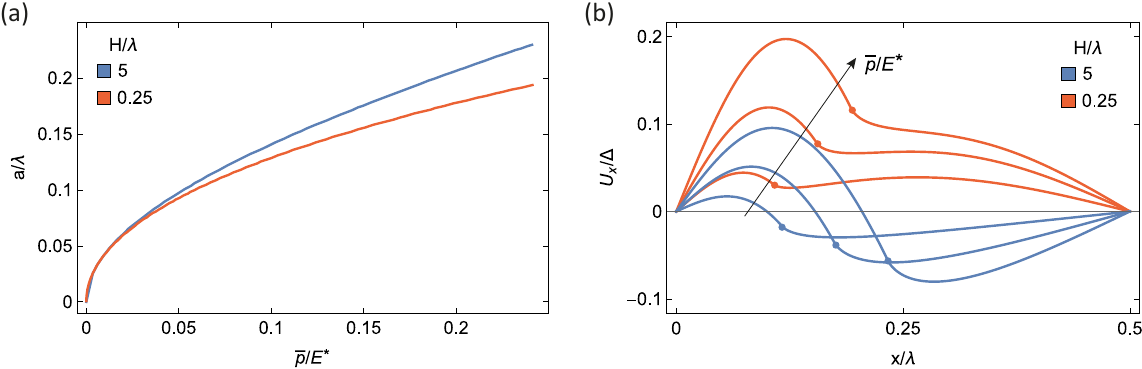}
\end{center}
\caption{\rev{The effect of the layer thickness on the loading phase in frictionless interfacial condition for the FE nonlinear contact model (NL)}. (a) The dimensionless contact area $a/\lambda$ as a function of the dimensionless applied pressure $\bar{p}/E^*$. (b) The dimensionless horizontal displacement $U_\mathrm{x}/\Delta$ versus the dimensionless contact position $x/\lambda$ (circles indicate the contact edges). Results are given for dimensionless external pressure values $\Bar{p}/E^*=0.07, 0.15, 0.25$. Curves refer to $H/\lambda=5$ and $0.25$ corresponding to unconfined and confined contact conditions, respectively, and are obtained for $\Delta/\lambda=0.18$.}
\label{fig9}%
\end{figure}

Fig. \ref{fig9} shows the effect of the mechanical confinement on the frictionless normal indentation \rev{in the framework of the nonlinear model (NL)}. In Fig. \ref{fig9}a, the dimensionless contact area $a/\lambda$ is plotted against the dimensionless applied mean pressure $\bar{p}/E^*$, while Fig. \ref{fig9}b shows the horizontal displacement field on the solid surface. Results are given for two values of the layer thickness $H/\lambda=5, 0.25$, corresponding to the half-plane and confined finite thickness conditions, respectively. 
Mechanical confinement leads to a stiffening of the contact behaviour, resulting in a smaller contact area for a given external pressure compared to the unconfined case\rev{, as qualitatively expected even in linear elastic approximation \cite{menga2016,Menga2018}}. More importantly, confinement amplifies horizontal displacements and modifies the nonlinear coupling between in-plane and out-of-plane displacements. Indeed, as depicted in Fig. \ref{fig9}b, confinement causes outward (positive) horizontal displacements to extend across the entire solid interface.

\begin{figure}[!ht]
\begin{center}
\includegraphics[width=\textwidth]{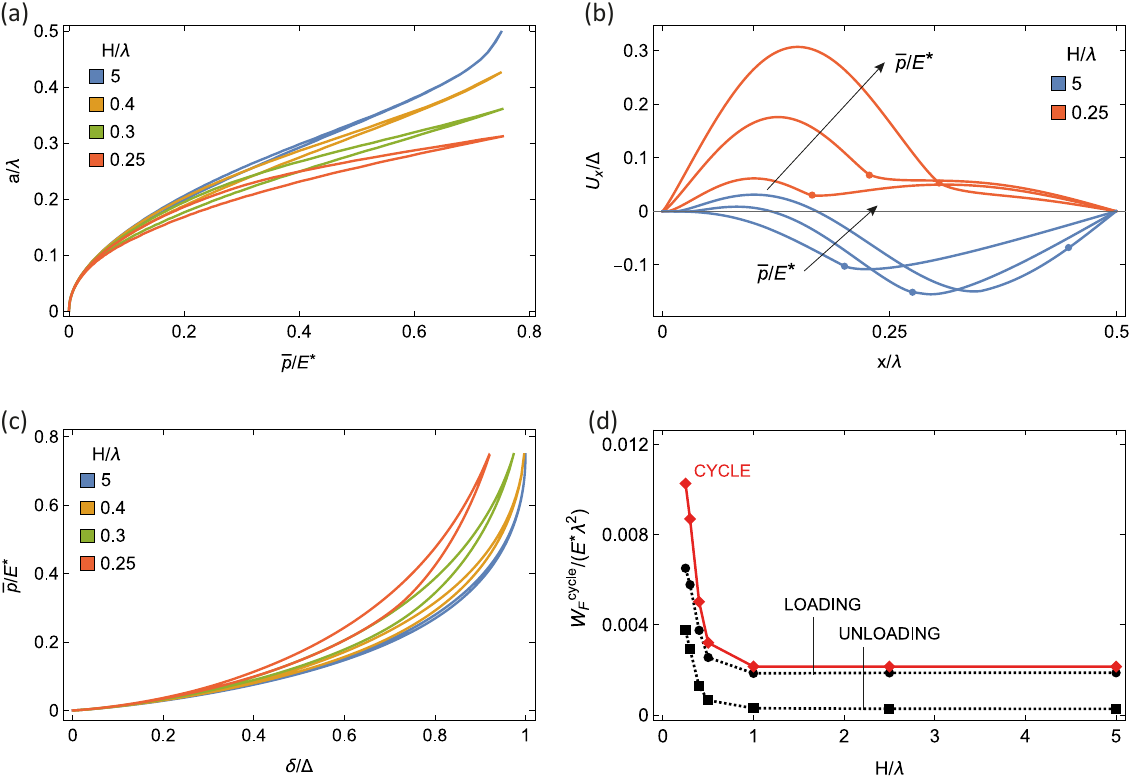}
\end{center}
\caption{\rev{The effect of the layer thickness on the normal loading-unloading cycle in frictional interfacial conditions for the FE nonlinear contact model (NL)}. (a) The dimensionless contact area $a/\lambda$ as a function of the dimensionless applied pressure $\Bar{p}/E^*$. (b) The dimensionless horizontal displacement field $U_\textrm{x}/\Delta$ (circles indicate the contact edges) for dimensionless external pressure values $\Bar{p}/E^*=0.2, 0.4, 0.75$.
(c) The dimensionless applied pressure $\Bar{p}/E^*$ as a function of the dimensionless penetration $\delta/\Delta$
(d) The dimensionless energy dissipated during a full indentation cycle $W_\textrm{F}^{\textrm{cycle}}/(E^*\lambda^2)$ as a function of $H/\lambda$ (red line); loading and unloading contributions (dashed lines) are also shown.
Results refer to $\Bar{p}_{\textrm{max}}/E^*=0.75$, $\tau_{\textrm{max}}/E^*=0.076$, and $\Delta/\lambda=0.18$.}%
\label{fig10}%
\end{figure}

Finally, Fig. \ref{fig10} illustrates the combined effect of friction and confinement \rev{on the contact response for the nonlinear model (NL)}. Figures \ref{fig10}a and \ref{fig10}b show the dimensionless applied pressure $\Bar{p}/E^*$ and the dimensionless contact area $a/\lambda$ as functions of the dimensionless penetration depth $\delta/\Delta$ and $\Bar{p}/E^*$, respectively, for different values of the dimensionless layer thickness $H/\lambda$. Figure \ref{fig10}c, on the other hand, plots the frictional energy dissipation as a function of the ratio $H/\lambda$, highlighting the contribution of each phase.
Confined contact is associated with higher frictional dissipation. Although a thinner layer results in a stiffer behaviour and a reduced contact area, slip affects a wider region compared to half-plane scenario due to the significantly enhanced normal-tangential coupling. 
Notably, a plateau in the dissipated energy is observed for $H\geq \lambda$. This outcome agrees with linear theory results, which predict that the solid is perturbed within a region with a principal dimension comparable to the contact area. In our case, since under full contact condition $2a=\lambda$, the effect of confinement becomes negligible as long as $H\geq \lambda$.

The amplification of nonlinear coupling between normal and tangential \rev{local actions} in confined systems, as observed in our simulations, is particularly relevant for designing soft materials used in biomedical devices. For example, thin elastomeric tactile sensors have shown thickness-dependent performances \cite{Zhu2022} whose optimization can benefit from a deeper understanding of how nonlinearity affects confined contact behaviour. Similarly, transdermal drug delivery systems based on reusable adhesive thin patches might benefit from enhanced toughness induced by loading-unloading hysteresis loop \cite{Zhou2023}.

\section{Conclusions}

This study focuses on the effects of geometric and material nonlinearity on the contact mechanics between a wavy rigid indenter and a flat deformable substrate, using a finite element model developed \rev{in the framework of finite strains/displacements and neo-Hookean hyperelasticity. The analysis of the contact response aims at highlighting the difference with respect to the commonly adopted linear elastic case; for this reason, and to align with real-world contact conditions, we also include interfacial friction between the indenter and the deformable solid. 
The finite slope of the deformed contact interface leads to an overall stiffer contact response (i.e., lower contact area under given normal load) compared to classical linear elasticity predictions, especially for high values of the indenter aspect ratio. The discrepancy between real nonlinear case and theoretical linear predictions is exacerbated in the presence of interfacial friction (as in most of practical application); indeed, in nonlinear analysis, frictional shear stresses at the interface also affect the overall normal contact behavior, introducing a nonlinear coupling between normal and tangential elastic fields mostly related to geometric nonlinearity. 
As a consequence, focusing on loading-unloading cycles, our analysis indicates that frictional interfacial slips occur even for incompressible semi-infinite solids, ultimately resulting in hysteresis and energy dissipation within the cycle. This result strongly differs from classical predictions (no hysteresis), showing that linear elasticity cannot be straightforwardly applied in the design of functional interfaces in real engineering systems, from elastomeric seals and tires \cite{lorenz2013} to biomedical devices \cite{Zhou2023} and tactile sensors \cite{Zhu2022}, where high friction levels, finite size and/or surface morphology may trigger nonlinear effects. Indeed, while loading-unloading hysteresis in linear analysis can occur only due to additional phenomena (e.g., adhesion, viscoelasticity, plasticity), we show that geometric and material nonlinearity might explain it in real frictional contact, especially when the pull-off force vanishes.
Moreover, our analysis indicates that mechanical confinement of thin solids (e.g., coating) enhances nonlinear effects, stiffening the contact and inducing higher hysteretic losses due to increased nonlinear coupling between normal and tangential actions. This challenges the applicability of the common half-plane assumption, further highlighting the importance of considering the finite dimensions and nonlinearity of real-world materials.}
Overall, our findings emphasize the need to incorporate nonlinear effects in theoretical and computational studies to improve the accuracy of contact predictions, enhance the understanding of experimental observations, and develop specific design methodologies for cutting-edge applications.

\section*{Authors contribution statement}

M.C.: Numerical modelling; Investigation; Methodology; Preparation of the figures; Writing--Original-Draft.
G.V.: Conceptualization; Numerical modelling; Methodology; Writing--Original-Draft; Writing--Review and editing.
L.A.: Conceptualization; Methodology; Writing--Review and editing; Supervision.
N.M.: Conceptualization; Methodology; Writing--Review and editing; Supervision.

\section*{Data availability}
The datasets used and/or analyzed during the current study are available from the corresponding author on reasonable request.

\section*{Acknowledgements}
This work was supported by the Italian Ministry of University
and Research under the Programme ‘‘Department of Excellence’’ Legge
232/2016 (Grant No. CUP - D93C23000100001) and by the European Union - NextGenerationEU through the Italian Ministry of University and Research under the programs: PRIN2022 (Projects of Relevant National Interest) grant nr. 2022SJ8HTC - ELectroactive gripper For mIcro-object maNipulation (ELFIN); PRIN2022 PNRR (Projects of Relevant National Interest) grant nr. P2022MAZHX - TRibological
modellIng for sustainaBle design Of induStrial friCtiOnal inteRfacEs
(TRIBOSCORE).\\

\end{document}